\documentclass[english,aps, prl, twocolumn]{revtex4}
\usepackage[T1]{fontenc}
\usepackage[latin9]{inputenc}
\usepackage{units}
\usepackage{amsmath}
\usepackage{graphicx}
\usepackage{amssymb}

\makeatletter

\providecommand{\tabularnewline}{\\}

\@ifundefined{textcolor}{}
{%
 \definecolor{BLACK}{gray}{0}
 \definecolor{WHITE}{gray}{1}
 \definecolor{RED}{rgb}{1,0,0}
 \definecolor{GREEN}{rgb}{0,1,0}
 \definecolor{BLUE}{rgb}{0,0,1}
 \definecolor{CYAN}{cmyk}{1,0,0,0}
 \definecolor{MAGENTA}{cmyk}{0,1,0,0}
 \definecolor{YELLOW}{cmyk}{0,0,1,0}
 }

\makeatother

\makeatother

\usepackage{babel}

\makeatother

\usepackage{babel}

\makeatother

\usepackage{babel}

\makeatother

\usepackage{babel}

\makeatother

\usepackage{babel}

\begin{document}

\title{Electron-phonon coupling and charge gap in spin-density-wave iron-pnictides
from quasiparticle relaxation dynamics }

\author{L. Stojchevska$^{1}$, P. Kusar$^{1}$, T. Mertelj$^{1}$, V.V. Kabanov$^{1}$,
X. Lin$^{2}$, G. H. Cao$^{2}$, Z. A. Xu$^{2}$ and D. Mihailovic$^{1}$}

\affiliation{$^{1}$Complex Matter Dept., Jozef Stefan Institute, Jamova 39, Ljubljana,
SI-1000, Ljubljana, Slovenia }

\affiliation{$^{2}$Department of Physics, Zhejiang University, Hangzhou 310027,
People\textquoteright{}s Republic of China}

\date{\today}
\begin{abstract}
We investigate the quasiparticle relaxation and low-energy electronic
structure in undoped SrFe$_{2}$As$_{2}$ exhibiting spin-density
wave (SDW) ordering using optical pump-probe femtosecond spectroscopy.
A remarkable critical slowing down of the quasiparticle relaxation
dynamics at the SDW transition temperature $T_{SDW}=200\mbox{K}$
is observed. From temperature dependence of the transient reflectivity
amplitude we determine the SDW-state charge gap magnitude, $2\Delta_{\mathrm{SDW}}/k_{B}T_{\mathrm{SDW}}=7.2\pm1$.
The second moment of the Eliashberg function, $\lambda\left\langle \mbox{\ensuremath{\left(\hbar\omega\right)}}^{2}\right\rangle =110\pm10$
meV$^{2}$, determined from the relaxation time above {\normalsize $T_{SDW}$,}
is similar to SmFeAsO and BaFe$_{2}$As$_{2}$ indicating a rather
small electron phonon coupling constant unless the electron-phonon
spectral function ($\alpha^{2}F\left(\omega\right)$) is strongly
enhanced in the low-energy phonon region.
\end{abstract}
\maketitle
The discovery of high-temperature superconductivity in iron-based
pnictides \cite{KamiharaKamihara2006,kamiharaWatanabe2008,RenChe2008}
has attracted a great deal of attention recently. The question of
the relative importance of the lattice and spin degrees of freedom
for the superconducting pairing interaction becomes immediately apparent,
since the superconductivity appears upon doping the parent materials%
\footnote{For a recent review see ref. \onlinecite{IshidaNakai2009}.%
} which show the spin-density wave (SDW) ground state. Understanding
the parent SDW compounds from the point of electron-phonon and not
only the spin-spin and spin-charge interactions is therefore beneficial
for understanding the nature of the superconducting coupling in the
doped compounds.

Time resolved spectroscopy has been very instrumental in elucidating
the nature of the electronic excitations in superconductors, particularly
cuprates \cite{StevensSmith1997,KabanovDemsar99,DemsarPodobnik1999,KaindlWoerner2000,DvorsekKabanov2002,SegreGedik2002,SchneiderDemsar2002,DemsarAveritt2003,GedikOrenstein2003,GedikBlake2004,KusarDemsar2005,KaindlCarnahan2005,KabanovDemsar2005,BianchiChen2005,SchneiderOnellion2005}
and recently also iron pnictides\cite{MerteljKabanov2009prl,MerteljKabanov2009jsnm,ChiaTalbayev2008,TorchinskyChen2009,MerteljKusar2010}.
Moreover, the relaxation kinetics can give us valuable information
on the electronic structure\cite{KabanovDemsar99} and electron-phonon
coupling\cite{KabanovAlexandrov2008}.

In this work we present a time-resolved femtosecond spectroscopy study
of SrFe$_{2}$As$_{2}$ in the normal and the SDW state. From the
photo-excited carrier relaxation dynamics we determine the electron-phonon
coupling parameters and the charge gap magnitude. We compare the results
with recent data\cite{MerteljKusar2010} in SmFeAsO and find that
they are similar both in the SDW and normal state with some minor
differences in the magnitude of the response at high temperatures. 

Optical experiments were performed using the standard pump-probe technique,
with 50 fs optical pulses from a 250-kHz Ti:Al$_{2}$O$_{3}$ regenerative
amplifier seeded with an Ti:Al$_{2}$O$_{3}$ oscillator. We used
the pump photons with doubled ($\hbar\omega_{\mathrm{P}}=3.1$ eV)
photon energy and the probe photons with 1.55 eV photon energy. The
pump and probe polarizations were perpendicular to each other and
oriented with respect to the crystals to obtain the maximum amplitude
of the response at low temperatures. The pump and probe beam diameters
were determined by measuring the transmittance of calibrated pinholes
mounted at the sample position\cite{KusarKabanov2008}. Single crystals
of SrFe$_{2}$As$_{2}$ were prepared by the self-flux method.\cite{LiLuo2009}
For optical measurements the cleaved crystals were glued on a Cu plate
mounted in an optical liquid-He flow cryostat.

\begin{figure}[tbh]
\begin{centering}
\includegraphics[angle=-90,width=0.3\textwidth]{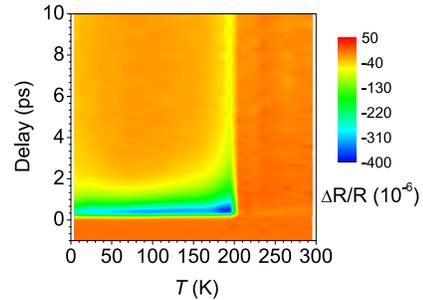} 
\par\end{centering}

\caption{(Color online) $\Delta R/R$ transients as a function of temperature
at pump-fluence 17 $\mu$J/cm$^{2}$. A divergent relaxation time
at $T_{\mathrm{SDW}}=200$K is clearly seen.}

\label{fig:fig-DR-2D} 
\end{figure}

\begin{figure}[tbh]
\begin{centering}
\includegraphics[width=0.45\textwidth]{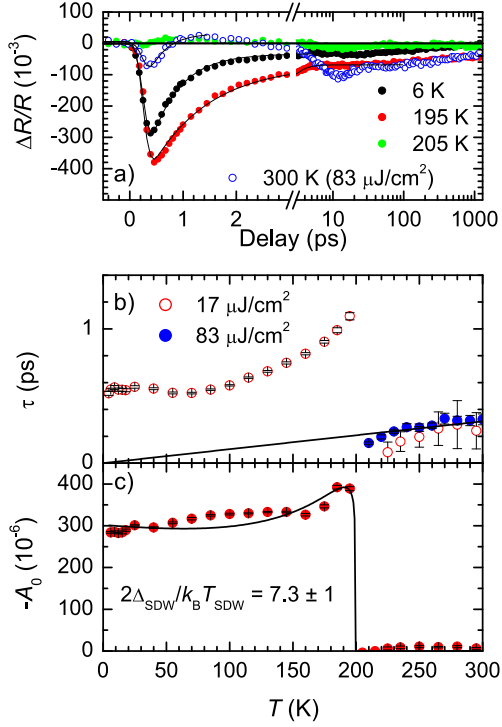} 
\par\end{centering}

\caption{(Color online) $\Delta R/R$ transients at representative temperatures
with single-exponential decay fits (a). The relaxation time at two
pump fluences (b) and amplitude at $\mathcal{F}=17$ $\mu$J/cm$^{2}$
(c) as functions of temperature. The black solid line in (b) is fit
of equation (\ref{eq:TauPoorHigh}) to $\tau$ above 230K. The black
solid line in (c) represents the fit of equation (6) from \cite{KabanovDemsar99}. }

\label{fig:fig-AvsT} 
\end{figure}

In Fig. \ref{fig:fig-DR-2D} we plot the temperature dependence of
$\Delta R/R$ transients in SrFe$_{2}$As$_{2}$. Below $T_{\mathrm{SDW}}$
the transients are dominated by the initial single exponential relaxation
followed by a weak structure at around 10 ps (see Fig. \ref{fig:fig-AvsT}
(a)). At $T_{\mathrm{SDW}}$ a critical slowing down of relaxation
is observed in the form of a long lived relaxation which is following
the initial $\sim1$ ps exponential decay. Above $T_{\mathrm{SDW}}$
the amplitude of the initial sub-ps relaxation strongly drops and
the structure on a longer timescale becomes apparent. The behavior
is similar to SDW SmFeAsO\cite{MerteljKusar2010} with the exception
of the sub-ps relaxation amplitude above $T_{\mathrm{SDW}}$ being
smaller in SrFe$_{2}$As$_{2}$. 

\begin{figure}[tbh]
\begin{centering}
\includegraphics[width=0.4\textwidth]{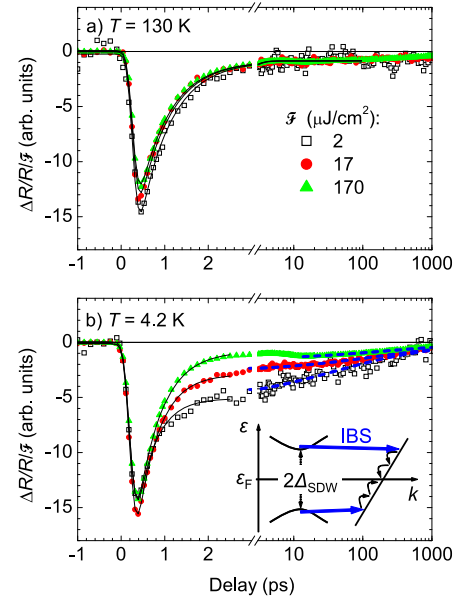} 
\par\end{centering}

\caption{(Color online) Fluence dependence of normalized $\Delta R/R$ transients
in the SDW state. The black thin lies are single exponential decay
fits and blue dashed lines the 1D diffusion\cite{MerteljOslak2009}
fits. In the inset to (b) the relaxation pathway via inter-band scattering,
that is discussed in text, is schematically shown.}

\label{fig:fig-DRvsF} 
\end{figure}

\begin{figure}[tbh]
\begin{centering}
\includegraphics[width=0.35\textwidth]{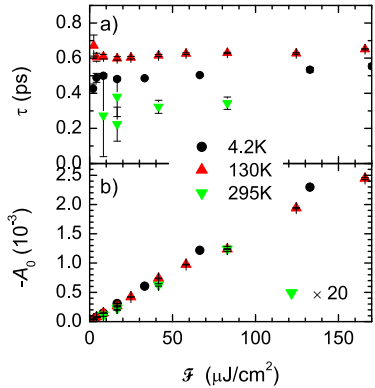} 
\par\end{centering}

\caption{(Color online) Fluence dependence of the $\Delta R/R$ transient amplitude
and relaxation time at different temperatures. }

\label{fig:fig-AvsF} 
\end{figure}

The amplitude of the initial sub-ps peak shows a minor departure from
the linear pump fluence ($\mathcal{F}$) dependence at the highest
$\mathcal{F}$ (see Figs \ref{fig:fig-DRvsF} and \ref{fig:fig-AvsF})
while the sub-ps relaxation time is virtually $\mathcal{F}$ independent
(see Fig. \ref{fig:fig-AvsF}). At low $T$ however, an additional
non-exponential slow relaxation component appears at the lowest fluence
(see Fig. \ref{fig:fig-DRvsF}) which can be (beyond 5-10 ps) attributed
to the heat diffusion out of the excitation volume\cite{MerteljOslak2009}
as indicated by fits in Fig. \ref{fig:fig-DRvsF}.


In a metal the photoexcited-quasiparticle relaxation time is governed
by transfer of energy from electronic degrees of freedom to the lattice
degrees of freedom. Recently the problem was solved analytically\cite{KabanovAlexandrov2008}.
In bad metals, above the Debye frequency ($\omega_{\mathrm{D}}),$
the relaxation time linearly depends on the temperature, $T$, where
the slope is determined by the inverse of the second moment of the
Eliashberg function $\lambda\langle\omega^{2}\rangle$:\cite{KabanovAlexandrov2008,GadermaierAlexandrov2009}

\begin{equation}
\tau=\frac{2\pi k_{\mathrm{B}}T}{3\hbar\lambda\langle\omega^{2}\rangle}.\label{eq:TauPoorHigh}\end{equation}
We find that above $\sim230$ K our $\tau$-data nicely follow the
predicted linear $T$ dependence (see Fig. \ref{fig:fig-AvsT} (b))
with $\lambda\langle\mbox{\ensuremath{\left(\hbar\omega\right)}}^{2}\rangle=110\pm10$meV$^{2}$.
The phonon spectrum of SrFe$_{2}$As$_{2}$ extends up to $\sim$40meV
with the acoustic phonon cutoff at $\sim10$meV.\cite{MittalZbiri2009}
We determine $\lambda\langle\mbox{\ensuremath{\left(\hbar\omega\right)}}^{2}\rangle/\lambda$
by assuming that the electron-phonon spectral function, $\alpha^{2}F\left(\omega\right)$,
has the phonon DOS shape\cite{BoeriDolgov2008,MittalZbiri2009} and
obtain $\lambda\langle\mbox{\ensuremath{\left(\hbar\omega\right)}}^{2}\rangle/\lambda\simeq430$
meV$^{2}$. This gives $\lambda\simeq0.25$, which is rather low to
explain the superconducting critical temperatures in doped compounds
within a standard single band BCS model.\cite{BoeriDolgov2008} If
however, $\alpha^{2}F\left(\omega\right)$ is for some reason enhanced
in the low-energy phonons region, $\lambda$ could easily reach significantly
higher values.%

Below $T_{\mathrm{SDW}}$ the increasing amplitude of the sub-ps transients
indicates the appearance of a bottleneck in the photo-excited electron
relaxation. The bottleneck is associated with opening of a $T$-dependent
gap\cite{KabanovDemsar99} due to SDW formation resulting in the Fermi
surface reconstruction\cite{sebastianGillett2008,AnalytisMcDonald2009}.
We use equation (6) from Kabanov \textit{et al.}\cite{KabanovDemsar99},
which describes the photo-excited change in quasiparticle density
in the presence of a temperature dependent gap, to fit the amplitude
below $T_{\mathrm{SDW}}=200$ K. Using a single SDW gap energy with
the BCS temperature dependence and $2\Delta_{\mathrm{SDW}}/k_{\mathrm{B}}T_{\mathrm{SDW}}\simeq7.2\pm1$
results in a rather good fit to the amplitude temperature dependence
(see Fig. \ref{fig:fig-AvsT} (c)). The observed gap magnitude is
close to the magnitude of the largest of the two SDW gaps obtained
from the optical conductivity.\cite{HuDong2008} Since the inter-band
scattering is strong, as argued below, the smallest of the gaps should
present the bottleneck for the hot carrier energy relaxation so our
data do not confirm the existence of the smaller gap suggested by
Hu \textit{et al.}\cite{HuDong2008}.

Similarly to the case of SmFeAsO\cite{MerteljKusar2010} we observe
no decrease of the relaxation time with $\mathcal{F}$ as predicted
by Kabanov \textit{et al.}\cite{KabanovDemsar99}. We also observe
no divergence of the relaxation time with decreasing $T$ such as
in heavy fermion SDW UNiGa$_{5}$,\cite{ChiaZhu2006} which, similar
to as SrFe$_{2}$As$_{2}$,\cite{ChenLi2008} remains metallic\cite{MorenoBauer2005}
upon the SDW gap opening. We can rule out the ballistic hot electron
escape as a source of the low-$T$ relaxation-time divergence cutoff,\cite{DemsarKabanov2009}
due to the relatively high resistivity of SrFe$_{2}$As$_{2}$\cite{ChenLi2008}
and large optical penetration depth of $\sim60$ nm at $\hbar\omega_{\mathrm{probe}}=1.55$
eV.%
\footnote{The optical penetration depth estimation is based on the ellipsometry
data in LaFeAsO\cite{BorisKovaleva2009} and similarity of the optical
conductivity in SrFe$_{2}$As$_{2}$\cite{HuDong2008} to that of
LaFeAsO.%
} However, in iron pnictides, due to the presence of ungapped electronic
bands below $T_{\mathrm{SDW}}$, the energy relaxation is not limited
by the anharmonic energy transfer from the high frequency to the low
frequency phonons, but rather by the inter-band scattering (IBS) from
the states at the edge of the SDW gap to the states in ungapped band(s)
with energies $\epsilon-\epsilon_{\mathrm{F}}\gg k_{\mathrm{B}}T$
(see inset to Fig. \ref{fig:fig-DRvsF}(b)). Such scattering can be
enhanced by the presence of impurities, which may explain the difference
between the higher residual resistivity SrFe$_{2}$As$_{2}$ and lower
residual resistivity UNiGa$_{5}$. 

The absence of multiple relaxation components (apart from diffusion)
from the $\Delta R/R$ transients together with IBS imply that mainly
the hot carriers from the SDW gaped electronic bands contribute to
the photoinduced reflectivity transients. This does not hinder the
determination of $\lambda\langle\left(\hbar\omega\right){}^{2}\rangle$
at high temperatures, since the IBS is essentially a momentum scattering,
which becomes faster than the energy relaxation rate at high temperatures. 

By means of the femtosecond optical pump-probe spectroscopy we determined
the second moment of the Eliashberg function in SDW SrFe$_{\mathrm{2}}$As$_{\mathrm{2}}$
and compared it to SmFeAsO and BaFe$_{2}$As$_{2}$. In all compounds
the values are similar (see Table I) suggesting moderate values of
the electron phonon coupling constant $\lambda$ as suggested by the
linear response theory,\cite{BoeriDolgov2008} unless $\alpha^{2}F\left(\omega\right)$
is strongly enhanced in the low-energy phonons region. 

Below $T_{\mathrm{SDW}}$ the temperature dependence of the relaxation
indicates the appearance of a quasi-particle relaxation bottleneck
due to opening of a single charge gap at $T\mathrm{_{SDW}}$ with
a BCS-like temperature dependence similar to that observed\cite{MerteljKusar2010}
in SmFeAsO and inferred from the recently published data\cite{ChiaTalbayev2010}
in BaFe$_{2}$As$_{2}$.%
\begin{table}[h]
\begin{centering}
\bigskip{}
\begin{tabular}{c|cccc}
\multicolumn{1}{c|}{} & \multicolumn{1}{c}{$\lambda\langle\mbox{\ensuremath{\left(\hbar\omega\right)}}^{2}\rangle$} & $\lambda\langle\mbox{\ensuremath{\left(\hbar\omega\right)}}^{2}\rangle/\lambda$  & $\lambda$ & $\nicefrac{2\Delta_{\mathrm{SDW}}}{k_{\mathrm{B}}T_{\mathrm{SDW}}}$\tabularnewline
 & (meV$^{2}$) & (meV$^{2}$) &  & \tabularnewline
\hline 
SrFe$_{\mathrm{2}}$As$_{\mathrm{2}}$  & $110\pm10$  & $430$ & $\sim0.25$ & $7.2\pm1$\tabularnewline
SmFeAsO ref.\cite{MerteljKusar2010}  & $135\pm10$ & $770$ & $\sim0.18$ & $\sim5$\tabularnewline
BaFe$_{\mathrm{2}}$As$_{\mathrm{2}}$ %
\footnote{We used the data from the supplemental material of Chia \textit{et
al.} {[}\onlinecite{ChiaTalbayev2010}{]}. (http://prl.aps.org/supplemental/PRL/v104/i2/e027003).%
} & $\sim65$%
\footnote{Only data just above $T_{\mathrm{SDW}}\simeq130K$ is available so
the value might be underestimated.%
} & $430$ & $\sim0.15$ & $4.7\pm1.6$\tabularnewline
\end{tabular}
\par\end{centering}

\begin{centering}
\caption{Electron phonon coupling parameters and SDW gap magnitudes in SDW
iron-pnictides. The experimental inelastic-neutron-scattering phonon
spectra from Mittal \textit{et al.} \cite{MittalZbiri2009,MittalSu2008}
and Osborn \textit{et al.} \cite{OsbornRosenkranz2009} were used
to estimate $\lambda\langle\mbox{\ensuremath{\left(\hbar\omega\right)}}^{2}\rangle/\lambda$
assuming $\alpha^{2}F\left(\omega\right)$ has the phonon DOS shape.}

\par\end{centering}

\centering{}\label{couplings}
\end{table}

\begin{acknowledgments}
This work has been supported by Slovenian Research Agency and the
National Science Foundation of China .
\end{acknowledgments}
\bibliography{biblio}

\end{document}